# How chromatin interactions shed light on interpreting non-coding genomic variants: opportunities and future directions

Yuheng Liang*[1], Sedigheh Abedini[1,2], Nona Farbehi[3], Hamid Alinejad-Rokny[1,]

1 UNSW BioMedical Machine Learning Lab (BML), School of Biomedical Engineering, UNSW Sydney, Sydney, NSW, 2052, Australia.

2 School of Biotechnology and Biomolecular Sciences, UNSW Sydney, Sydney, NSW, 2052, Australia.

3 School of Biomedical Engineering, UNSW Sydney, Sydney, NSW, 2052, Australia.

Correspondence should be address to: yuheng.liang@student.unsw.edu.au

Simple Summary: The non-coding region constitutes a significant portion of the human genome and plays a crucial role in the regulatory network. Investigating non-coding CNVs has thus emerged as one of the most important avenues in genomics research. Recent advancements in Hi-C data have provided three-dimensional mapping of chromatin interactions, offering new insights into overcoming the challenges associated with studying the effects of non-coding CNVs. In this review, we examine the importance of chromatin interactions in the interpretation of non-coding variants. Through a combination of literature review and in-house analyses, we identified non-coding CNVs that disrupt chromatin interactions, significantly influencing chromosomal interactions, regulatory networks, Topologically Associating Domains (TADs), single promoter-enhancer loops, compartmentalization, epigenetic modifications, and super-enhancers.

**Abstract:** Genomic variants, including copy number variants (CNVs) and genome-wide association study (GWAS) single nucleotide polymorphisms (SNPs), represent structural alterations that influence genomic diversity and disease susceptibility. While coding region variants have been extensively studied, non-coding and regulatory variants present significant challenges due to their potential impacts on gene regulation, which are often obscured by the complexity of the genome. Chromatin interactions, which organize the genome spatially and regulate gene expression through enhancer-promoter contacts, predominantly occur in non-coding regions. Notably, more than 90% of enhancers, crucial for gene regulation, reside in these non-coding regions, underscoring their importance in interpreting the regulatory effects of CNVs and GWAS-associated SNPs. In this study, we integrate chromatin interaction data with CNV and GWAS data to uncover the functional implications of non-coding variants. By leveraging this integrated approach, we provide new insights into how structural variants and disease-associated SNPs disrupt regulatory networks, advancing our understanding of genetic complexity. These findings offer potential avenues for personalized medicine by elucidating disease mechanisms and guiding therapeutic strategies tailored to individual genomic profiles. This research underscores the critical role of chromatin interactions in revealing the regulatory consequences of non-coding variants, bridging the gap between genetic variation and phenotypic outcomes.

# Introduction

Non-coding genomic variants, including non-coding copy number variants (CNVs) and single nucleotide polymorphisms (SNPs), particularly those uncovered in Genome-Wide Association Studies (GWAS), represent significant alterations in the genome. These variants frequently impact critical regulatory elements, such as enhancers, promoters, and insulators, which are essential for precise spatiotemporal control of gene expression. These variants can involve structural changes (i.e., deletions, duplications, and re-arrangements) within regions that constitute approximately 98% of the human genome [1]. This vast non-coding landscape, once considered "junk DNA," is now recognized as a key player in cellular function and organismal development [1, 2].

Despite their prevalence and potential biological significance, non-coding variants have been historically underexplored compared to coding variants. This gap in research stems from the complexity of elucidating their functional roles in the intricate network of gene regulation [3, 4]. Nonetheless, the growing evidence of their impact on phenotypic diversity and disease susceptibility is driving deeper investigations into these important genomic elements [5].

The regulation of gene expression in eukaryotic cells is a multifaceted process governed not only by the linear DNA sequence but also by the three-dimensional organization of the genome within the nucleus [6]. This spatial organization is largely mediated by chromatin interactions, which bring distant genomic loci into close physical proximity. These interactions are crucial for the proper functioning of regulatory elements, particularly in facilitating communication between enhancers and their target gene promoters [7]. Chromatin interactions predominantly occur in non-coding regions, where over 90% of enhancers are located [8]. Enhancers serve as critical modulators of gene transcription, often acting over long genomic distances. Disruptions of enhancer-promoter interactions, potentially caused by non-coding variants, can lead to aberrant gene expression patterns, contributing to various cellular disruptions and disease phenotypes [9].

We have recently demonstrated the significance of genomic variants, including single nucleoide polimorphisms in cancer [10-12] and CNVs, in genetic diseases [13-16]. However, interpreting non-coding variants remains a significant challenge. The study of chromatin interactions offers a promising approach to elucidating the regulatory impact of non-coding variants [17, 18]. By mapping the three-dimensional genome, researchers can better understand how non-coding variants disrupt or modify the interactions between regulatory elements and their target genes. This understanding sheds light on gene regulation mechanisms and the potential consequences of genomic alterations. [19]. For example, non-coding variants may interfere with an enhancer's ability to properly interact with a promoter, potentially altering gene expression and leading to disease phenotypes [4]. Advanced high-resolution chromatin interaction mapping techniques, such as Hi-C and ChIA-PET, have revolutionized our ability to visualize and analyze these interactions. These methods provide detailed insights into such interactions and their potential alterations across different cell types and physiological conditions, revealing the dynamic nature of genome organization [20].

The integration of chromatin interaction data with the analysis of non-coding variants offers a more comprehensive understanding of the genome's regulatory architecture, enabling the identification of functionally relevant non-coding variants that play key roles in gene regulation and disease [21, 22]. By examining these regulatory interactions, researchers can uncover previously hidden genetic factors contributing to disease susceptibility, opening new avenues for therapeutic intervention [23]. Insights into how non-coding variants reshape chromatin structure may inform the development of therapies aimed at restoring normal gene expression [24]. Additionally, studying chromatin interactions in this context has broader implications for understanding genome evolution, as many non-coding regulatory elements are conserved across species. This conservation offers insights into adaptive changes and the maintenance of genomic integrity [25].

In this study, we explore how chromatin interaction data can enhance the interpretation of non-coding variants by providing insights into their regulatory roles. By integrating through computational analyses chromatin interactions with genomic data, we aim to elucidate the complex relationship of non-coding variants with gene regulation and chromatin architecture and their functional consequences.

## Method & data collection

**Preparation of Hi-C Libraries**

We utilised in-house Hi-C data from a muscular cell line to perform analysis. Two biological replicates were available for each cell line, ensuring reproducibility of the observed chromatin interactions. The Hi-C data were processed and aligned using HiC-Pro v2.11 [26] with default parameters, employing the hg19 genome build for mapping at a 5 kb resolution. This resolution balances fine-scale interaction detection with robust statistical power.

To identify statistically significant interactions, we employed MaxHiC [27] and MHiC [28]. We applied stringent criteria to define significant interactions: a P-value < 0.01, a read count ≥ 10, and distances between interacting fragments ranging from 5 kb to 10 Mb. Hi-C interactions were then annotated with coding and non-coding genes from our combined and curated gene list sourced from FAMTOM5 [29], ENSEMBL [30] and ENCODE [31]. We required a minimum of 10% overlap between the gene promoter and the Hi-C fragments for inclusion in the analysis, ensuring that the detected interactions were likely to be functionally relevant to the genes of interest.

**Genomic Variants**

We collected data on two types of genomic variants: GWAS SNPs and CNVs. GWAS SNPs were downloaded from two comprehensive databases: the GWAS Catalog (https://www.ebi.ac.uk/gwas/docs/file-downloads) [32, 33] and GWASdb v2 (http://jjwanglab.org/gwasdb) [34]. Neuromuscular-associated CNVs were collected from [35], comprising 60 deletions and 19 duplications. To ensure compatibility with our Hi-C data and current genomic annotations, all CNVs were lifted over from the hg19 to the hg38 genome build using the UCSC LiftOver tool (https://genome.ucsc.edu/cgi-bin/hgLiftOver) [36] and the Genebe Liftover tool (https://genebe.net/tools/liftover)[37]. Two out of the 79 CNVs mapped to multiple regions in the new build and were excluded from further analysis. The list of CNV regions is provided in **Table 1**.

**Table 1.** list of Neuromuscular-associated CNVs obtained from [35].

| chrom_hg38 | start_hg38 | end_hg38 | cnv_type | loci |
|---|---|---|---|---|
| chr2 | 32063610 | 32224112 | del | 2p22.3p22.3 |
| chr2 | 32087416 | 32175421 | del | 2p22.3p22.3 |
| chr2 | 74073548 | 74682058 | dup | 2p13.2p13.2 |
| chr2 | 151609804 | 151725549 | del | 2q23.3q23.3 |
| chr2 | 151663519 | 152335728 | dup | 2q23.3q23.3 |
| chr2 | 163593644 | 173051809 | dup | 2q24.2q31.1 |
| chr2 | 178530240 | 178578185 | del | 2q31.2q31.2 |
| chr2 | 178692208 | 178736793 | dup | 2q31.2q31.2 |
| chr2 | 237358411 | 237360280 | dup | 2q37.3q37.3 |
| chr2 | 237364304 | 237372349 | del | 2q37.3q37.3 |
| chr2 | 237574321 | 242095027 | del | 2q37q37 |
| chr3 | 4516493 | 4710557 | del | 3p26.1p26.1 |
| chr6 | 5216531 | 5545368 | del | 6p25.1p25.1 |
| chr6 | 5368562 | 5369231 | del | 6p25.1p25.1 |
| chr6 | 152364787 | 152369675 | del | 6q25.2q25.2 |
| chr6 | 161973201 | 161973430 | dup | 6q26q26 |
| chr6 | 161973201 | 162201287 | del | 6q26q26 |
| chr6 | 161973201 | 162054212 | del | 6q26q26 |
| chr6 | 162201048 | 162262738 | del | 6q26q26 |
| chr6 | 162201048 | 162201287 | del | 6q26q26 |
| chr6 | 162262429 | 162262738 | del | 6q26q26 |
| chr7 | 66638252 | 66643229 | del | 7q11.21q11.21 |
| chr7 | 93101303 | 95324501 | del | 7q21.2q21.3 |
| chr7 | 94395074 | 94911475 | del | 7q21.3q21.3 |
| chr7 | 94598745 | 94600829 | del | 7q21.3q21.3 |
| chr8 | 38249701 | 38255425 | del | 8p11.23p11.23 |
| chr11 | 34957403 | 34960560 | del | 11p13p13 |
| chr12 | 57619857 | 57630290 | del | 12q13.3q13.3 |
| chr13 | 23320636 | 23320760 | del | 13q12.12q12.12 |
| chr13 | 110149111 | 110357616 | dup | 13q34q34 |
| chr14 | 35595609 | 39125228 | del | 14q13.2q21.1 |
| chr14 | 36513111 | 36875004 | del | 14q13.3q13.3 |
| chr14 | 54902294 | 54902669 | del | 14q22.1q22.1 |
| chr14 | 87933766 | 87950035 | del | 14q31.3q31.3 |
| chr14 | 87934878 | 87950658 | del | 14q31.3q31.3 |
| chr15 | 23439583 | 28299576 | del | 15q11.2q13.1 |
| chr15 | 42384459 | 42392727 | del | 15q15.1q15.1 |
| chr15 | 44570425 | 44575017 | del | 15q21.1q21.1 |
| chr15 | 44570468 | 44575076 | del | 15q21.1q21.1 |
| chr15 | 44570468 | 44575076 | del | 15q21.1q21.1 |
| chr15 | 44570468 | 44575076 | del | 15q21.1q21.1 |
| chr15 | 44570468 | 44575076 | del | 15q21.1q21.1 |
| chr15 | 44585655 | 44604204 | dup | 15q21.1q21.1 |
| chr15 | 44595293 | 44596925 | del | 15q21.1q21.1 |
| chr16 | 14822853 | 16221747 | dup | 16p13.11p13.11 |
| chr16 | 28391733 | 29365070 | del | 16p11.2p11.2 |
| chr16 | 29457619 | 30188525 | del | 16p11.2p11.2 |
| chr16 | 89507631 | 89509073 | del | 16q24.3q24.3 |
| chr16 | 89529482 | 89554569 | del | 16q24.3q24.3 |
| chr16 | 89529482 | 89554569 | del | 16q24.3q24.3 |
| chr16 | 89556827 | 89557033 | del | 16q24.3q24.3 |
| chr17 | 14191988 | 15574208 | del | 17p12p12 |
| chr17 | 14191988 | 15574208 | dup | 17p12p12 |
| chr17 | 14191988 | 15574208 | dup | 17p12p12 |
| chr17 | 50392397 | 50619826 | dup | 17q21.33q21.33 |
| chr17 | 80118139 | 80118367 | del | 17q25.3q25.3 |
| chr18 | 1168409 | 14764090 | del | 18p11.32p11.21 |
| chr18 | 2771470 | 3277980 | del | 18p11.31p11.31 |
| chr18 | 9708321 | 12884299 | del | 18p11.22p11.21 |
| chr19 | 13224557 | 13235834 | del | 19p13.2p13.2 |
| chr19 | 13334391 | 13335915 | del | 19p13.2p13.2 |
| chr19 | 13359799 | 13365469 | del | 19p13.2p13.2 |
| chr20 | 2835336 | 3974384 | del | 20p13p13 |
| chr22 | 18429207 | 18659564 | dup | 22q11.21q11.21 |
| chr22 | 18429207 | 18659564 | del | 22q11.21q11.21 |
| chr22 | 18906447 | 20319895 | dup | 22q11.2q11.2 |
| chr22 | 20043354 | 20063442 | del | 22q11.21q11.21 |
| chr22 | 33274445 | 33337828 | dup | 22q12.3q12.3 |
| chrX | 30833316 | 31267927 | del | Xp21.1p21.1 |
| chrX | 31169370 | 31627909 | del | Xp21.1p21.1 |
| chrX | 31478153 | 31932228 | dup | Xp21.1p21.1 |
| chrX | 31657098 | 32155908 | dup | Xp21.1p21.1 |
| chrX | 31819974 | 31968514 | del | Xp21.1p21.1 |
| chrX | 31875190 | 31968514 | del | Xp21.1p21.1 |
| chrX | 32411751 | 32699293 | del | Xp21.1p21.1 |
| chrX | 32438241 | 32644313 | del | Xp21.1p21.1 |
| chrX | 150592506 | 150829533 | del | Xq28q28 |

For downstream Hi-C analysis, matrices generated by HiC-Pro were converted into both H5 and COOL formats using hicExplorer [38-40]. Topologically Associating Domains (TADs), fundamental units of chromatin organization, were identified along with their boundaries, regions, and TAD separation scores using hicExplorer. A false discovery rate (FDR) threshold of < 0.05 was applied to a identify statistically significant TAD boundaries and account for multiple comparisons.. In addition to TADs, chromatin loops—long-range interactions often involved in gene regulation—were identified using hicExplorer.. To further characterize the higher-order chromatin architecture, we employed HiTC [41], an R-based toolset compatible with HiC-Pro outputs, to detect A/B compartments, which correspond to transcriptionally active and inactive regions, respectively.

To assess the regulatory impact of GWAS-identified single nucleotide polymorphisms (SNPs) on chromatin architecture, we performed a comparative analysis by examining the overlap between GWAS SNPs and the TAD and chromatin loops identified in our previous analyses. Additionally, we quantified the number of GWAS SNPs within each A/B compartment to evaluate the effects of chromatin interactions across different genomic regions. All analyses and visualizations, including statistical comparisons and graphical representations, were conducted using R scripts.

Hi-C interactions and associated genomic features were visualized using hicExplorer [38-40] and pyGenomeTracks [42]. We generated multi-panel diagrams incorporating TAD regions, chromatin loops, compartments, ChIP-Seq data, FANTOM promoter and enhancer signals, and GWAS SNPs linked to muscular disease phenotypes. This approach enables a holistic view of the regulatory landscape in the context of three-dimensional genome organization.

All code and configuration files used in the analysis are available at https://github.com/jade0530/hic_review. To further investigate the epigenetic context of CNVs relevant to the neuromuscular disease phenotype, we generated diagrams integrating H3K4me1 and H3K27ac enhancer signals. Additionally, we visualized chromatin interaction profiles for significant GWAS SNPs linked to neuromuscular disease phenotype using the WashU genome browser.

## Results and discussion

### Alterations in chromatin interaction maps

The impact of non-coding variants on chromatin interaction maps has been increasingly understood through high-resolution chromosome conformation capture techniques, particularly Hi-C. These techniques have demonstrated that CNVs and GWAS SNPs can induce significant modifications in the 3D genomic landscape. Chromatin interactions, repsresented through contact maps protray chromsomal links, which facilitate understanding of genomic abnoramlity (Figure 1A). Recent studies have shown that duplications can create novel regulatory domains by forming ectopic chromatin maps (Figure 1B) that bring enhancers into proximity with unrelated genes, potentially leading to aberrant gene activation or repression [43-45] potentiall through disruption of topological domain boundaries [23, 45, 46]. On the other hand, deletions can disrupt regulatory domains by breaking enhancer-target gene communication, thereby silencing essential genes or altering their expression patterns via perturbation of long-range chromatin interactions [47, 48]. Such structural alterations often result in genome-wide reorganization of chromatin interaction maps.

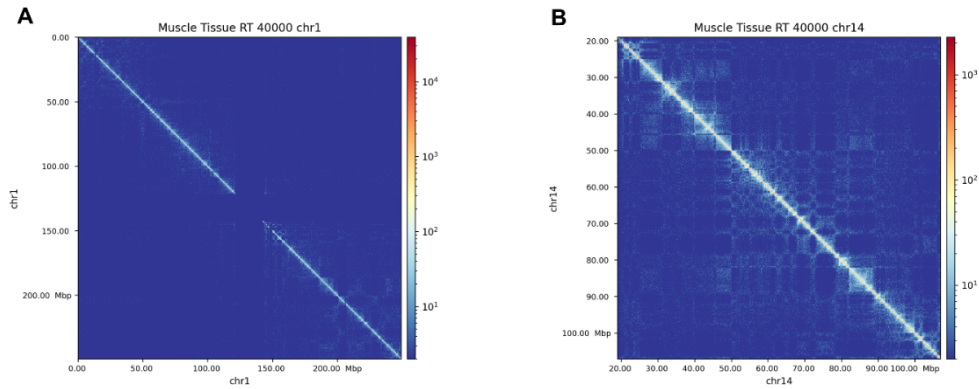

**Figure 1: Hi-C contact map.** Contact maps showing intra-chromosome interactions of chromosome 1 (left) and chromosome 14 (right), at 40k resolution. X and y-axis represents genomic locations, where heatmap scales depict contact probabilties under log1p transformation.

Large-scale CNVs, in particular, can affect TAD boundaries or structures, either merging distinct regulatory domains or creating new ones, which can have broad effects on gene regulation across large genomic regions. These structural changes can span multiple regulatory domains, leading to complex alterations in nuclear architecture [45, 49]. Additionally, these variants can alter the distribution of active (A) and inactive (B) compartments, leading to widespread changes in chromatin accessibility and gene expression patterns [50, 51]. Recent advancements in single-cell Hi-C have further illuminated how CNVs can induce distinct patterns of chromatin reorganization in different cell types, revealing cell-type specific variations in chromatin architecture that underlie tissue-specific manifestation of disease phenotypes [52, 53].

Genomic variants can profoundly influence chromatin architecture, leading to substantial misregulation of gene expression. The deletion or duplication of key regulatory elements, particularly those that define TAD boundaries, can result in the merging of adjacent TADs or the creation of new regulatory interactions through modification of architectural protein binding sites [23, 44, 46]. These changes can alter gene expression patterns and contribute to various genetic disorders. Wang et al. [54] demonstrated that genomic variants affecting TAD boundaries can lead to ectopic enhancer-promoter interactions, causing abnormal gene activation and contributing to disease pathogenesis. Their study revealed that these variants may introduce new chromatin loops or disrupt existing ones, depending on the specific genomic context. Such chromatin reorganization often affected the expression of multiple genes, leading to developmental abnormalities and underscoring the complexity of CNV-induced regulatory changes.

These findings highlight the importance of considering the broader genomic context when evaluating the pathogenicity of the variants. The three-dimensional organization of chromatin serves as a crucial regulatory layer that, when perturbed, can lead to widespread transcriptional dysregulation. They also suggest that the impact of structural variations extends beyond simple gene dosage effects, emphasizing the need for a more nuanced understanding of genome organization in both normal development and disease states.

**Cascading effects on cellular functions**

The reorganization of gene regulatory networks by either CNVs or GWAS can have cascading effects on cellular functions, potentially contributing to genomic instability and the development of diseases such as cancer and neurodevelopmental disorders. Genomic variants, such as CNVs, can alter genome architecture by causing the duplication or deletion of key regulatory elements, thereby impairing the expression of genes involved in critical cellular processes [55]. Such changes in genome architecture can disrupt essential processes, including DNA replication, repair, and transcription, ultimately affecting cellular phenotypes and disease progression [56-58].

For instance, disruptions in the *SHH* locus are linked to holoprosencephaly, a severe developmental disoreder [59], while CNV-induced alterations in the 22q11.2 region are associated with the variable clinical manifestations of DiGeorge syndrome [60]. These examples illustrate how alterations in genomic structure can lead to diverse clinical outcomes, underscoring the complex relationship between genotype and phenotype in CNV-associated disorders.

Furthermore, studies have shown that the total burden of rare CNVs, especially deletions, is strongly associated with disease risk [55]. This association is mainly driven by CNVs at known genomic disorder regions, whose pleiotropic effects on common diseases are broader than previously anticipated.

**Identification of regulatory networks**

High-throughput chromosome conformation capture techniques, such as Hi-C, have been instrumental in identifying regulatory networks by mapping spatial interactions between enhancers, promoters, and other regulatory elements. As shown in Figure 2, chromosomal interacting regions align with histone enhancer markers H3K27ac and H3K4me1 from hippocampus and prefrontal cortex. These interactions are essential for understanding how genomic variants disrupt normal gene regulation by altering the three-dimensional organization of the genome (Figure 2). Recent studies have revealed that these interactions are dynamically regulated during cellular differentiation, with specific transcription factor binding patterns determining the stability and specificity of enhancer-promoter contacts [61, 62]. The integration of GWAS data with functional genomics has significantly advanced our understanding of cancer-associated risk-variants. These variants are frequently enriched within chromatin regulatory enhancer regions, critical for controlling gene expression. By mapping these variants to specific enhancer regions, researchers have identified key regulatory elements that may contribute to the onset and progression of various cancers [63]. Recent comprehensive Hi-C analyses across multiple human tissues have demonstrated that the majority of distal regulatory elements interact with their target promoters within TADs, establishing a fundamental principle of genomic organization [64, 65].

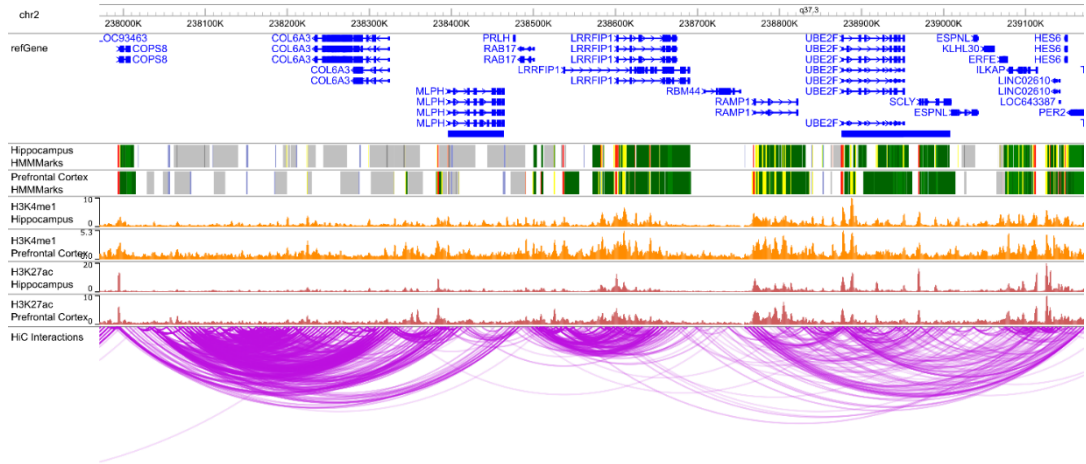

**Figure 2. Arc view of chromatin interacting regions.** Interacting regions that overlap with histone markers H3K27ac and H3K4me1 have the potential to function as regulatory regions.

The study of regulatory element duplications has provided insights into oncogene overexpression and cancer progression. Notable examples include the *MYC* and *PAX5* oncogenes, where enhancer duplications led to increased gene expression, subsequently promoting tumor growth [66, 67]. These findings emphasize the significance of dosage effects in gene regulation and their potential role in disease pathogenesis. Furthermore, S. Weinstock et al. employed CRISPR knockouts in CD4+ T cells, developing an innovative method to map gene regulatory networks. Their approach uncovered novel trans-regulatory relationships, emphasizing the pivotal role of disease-associated transcription factors in linking immune-related GWAS genes to critical signaling pathways [68]. . This work highlights the disruption of regulatory networks by genomic variants and their potential consequences for disease development. This has been further supported by recent single-cell studies showing that regulatory network plasticity plays a crucial role in cellular adaptation to perturbations [69]. Similarly, Tan et al. [70]. investigated the role of regulatory networks in heart failure by integrating genomic variant data with Hi-C analysis. They conducted epigenetic profiling in human heart tissue, identifying 47,321 potential enhancers and 3,897 differential acetylation sites associated with disrupted pathways in heart failure. Their study also revealed 1,680 histone acetylation Quantitative Trait Locus (haQTLs), some of which were linked to gene expression through long-range chromatin interactions identified by Hi-C [70]. Subsequent research has shown that these heart-specific regulatory networks are particularly sensitive to mechanical stress, with specific enhancer-promoter interactions being modulated by mechanical forces, providing new insights into the molecular basis of heart failure [71]. These researches underscore the importance of studying both genomic variants and regulatory networks to understand disease mechanisms.

**Disrupting topologically associating domain boundaries**

TADs are fundamental units of chromatin organization that create distinct regulatory environments within the genome. TAD boundaries act as insulators, ensuring that regulatory elements, such as, enhancers interact only with target genes within the same domain while preventing inappropriate cross-boundary interactions that could result in gene misregulation. Figure 3 depicts an example of TAD regions overlapping with CNVs and GWAS related SNPs. Mechanistic investigations have revealed that CCCTC-binding factor (CTCF), a zinc finger DNA-binding protein, and cohesin proteins are essential for TAD boundary formation. Studies demonstrate that CTCF depletion leads to a significant reduction in boundary strength and increased inter-TAD interactions [72, 73]. These domains are critical for maintaining proper gene expression, and disruptions to TAD boundaries have profound implications for human health and disease.

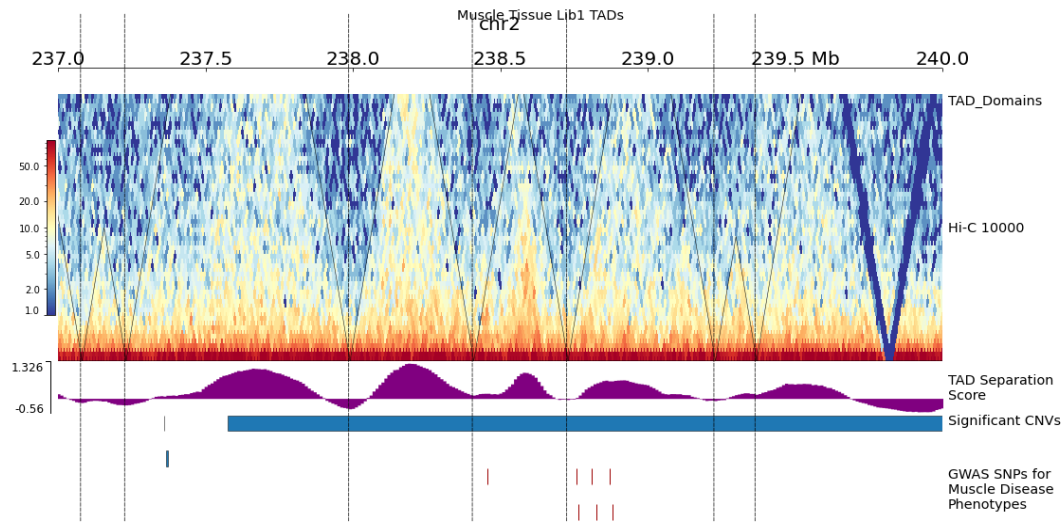

**Figure 3: Overview of TAD Regions.** Muscle disease CNVs were found to overlap with multiple TADs, suggesting potential disruptions in chromatin structure and gene regulation. Similarly, GWAS SNPs overlapped with three TADs, highlighting their potential functional role in modulating gene expression and contributing to disease mechanisms.

Genomic variants, particularly CNVs, that disrupt TAD boundaries can lead to improper enhancer-promoter interactions across different TADs, causing gene misregulation. This mechanism has been implicated in various developmental disorders and diseases. For instance, disruptions to TAD boundaries have been associated with limb malformations and congenital heart defects [74]. Advanced molecular analyses of these boundary disruptions have revealed that even subtle changes in boundary strength can lead to dosage-dependent effects on gene expression, with stronger boundary disruptions correlating with more severe phenotypes. A landmark study by Lupianez et al. [23] demonstrated that duplications and deletions affecting TAD boundaries could alter the regulatory landscape, causing the misexpression of key developmental genes and pathogenic phenotypes. Similarly, Franke et al. [75] provided further evidence linking CNVs disrupting TAD boundaries to congenital heart defects through aberrant gene activation. Recent research continues to underscore the importance of TAD boundaries in maintaining genomic integrity. For example, CNVs that disrupt TAD boundaries in neurodevelopmental disorders have been shown to cause significant gene misregulation, contributing to conditions such as autism and intellectual disability [76, 77]. These findings, based on advanced Hi-C and RNA-seq technologies, provide detailed insights into how structural variations impact chromatin organization and gene expression.

In our own analysis (Figure 3), we identified 4,098 muscle disease GWAS SNPs overlapping with 6,232 TAD domains identified through Hi-C data, compared to only 710 muscle disease GWAS SNPs overlapping with an equivalent number of randomly generated regions. Subsequent functional analysis of these overlapping regions revealed that SNPs located within 50kb of TAD boundaries were significantly more likely to affect gene expression through long-range regulatory interactions [64, 78, 79]. Additionally, integration with chromatin accessibility data showed that these boundary-proximal SNPs frequently coincided with tissue-specific regulatory elements [80]. This significant enrichment of disease-associated variants within TAD domains strongly supports the notion that TAD structures play a crucial role in the functional impact of genetic variations. It suggests that many disease-associated variants may exert their effects by altering chromatin organization and gene regulation rather than through direct changes to protein-coding sequences.

**Disrupting chromatin loops**

Chromatin loops are crucial for regulating gene expression by bringing distant genomic regions into close proximity, facilitating precise transcriptional control. Integrating genomic data with chromatin loops can provide valuable insights into interpreting variants, especially those that disrupt these loops and interfere with regulatory interactions, potentially leading to pathological outcomes, shown in Figure 4. Quantitative analysis of loop disruption events across multiple cancer types has revealed that almost pathogenic structural variants occur at loop anchor points, suggesting a mechanistic link between loop integrity and disease progression [81, 82]. For instance, duplications within a chromatin loop can introduce additional CTCF binding sites, which may alter the structure of the loops and result in gene misregulation [83-85].

CTCF plays a key role in the formation and maintenance of chromatin loops by binding to specific DNA sequences and facilitating interactions between loop anchors, often in conjunction with the cohesin complex [72, 73]. High-resolution structural studies have demonstrated that CTCF-mediated loops exhibit distinct stability patterns depending on the orientation of CTCF binding sites, with convergent sites showing significantly stronger loop formation than divergent arrangements [86, 87]. Changes in the number or arrangement of CTCF binding sites can perturb chromatin architecture, leading to abnormal gene expression patterns. Rao et al. [19] mapped the 3D structure of the human genome at kilobase resolution, demonstrating the critical function of CTCF and cohesin in organizing chromatin loops that regulate gene expression across various cell types and developmental stages. Similarly, Zhang et al. [88] highlighted dynamic promoter-enhancer interactions, illustrating how spatial genome organization governs gene expression during development.

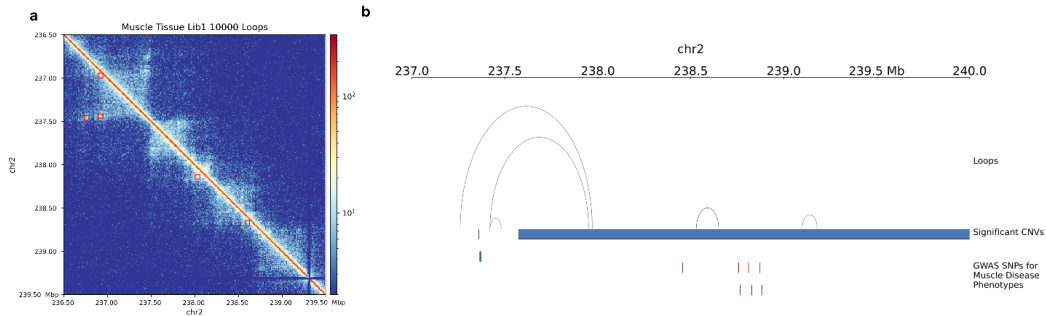

**Figure 4: Visualisations of Chromatin loops.** Identification of loops as marked red circle in contact map with 10k resolution (left) and mapped with muscle disease related CNVs and GWAS SNPs (right) in chromosome 2.

Tim et al. [89] demonstrated the critical role of chromatin loops as scaffolds for gene regulation during *Drosophila* embryogenesis, providing insights into the conservation of these regulatory mechanisms across species. Their investigation into the three-way *Scyl-Chrb* interaction, facilitated by introducing four genetic deletions, revealed that disrupting chromatin loops significantly alters the coordinated expression and cross-regulation of genes. Subsequent functional analyses revealed that loop disruption events trigger compensatory mechanisms involving alternative enhancer-promoter interactions, though these compensatory loops often exhibit reduced regulatory efficiency [90, 91]. These findings underscores the dynamic function of chromatin loops in fine-tuning gene expression across different developmental contexts and cellular states.

Genomic variants that disrupt chromatin loops pose a considerable threat to genomic stability and the precise regulation of genes. Understanding how these structural variations impact chromatin architecture offers valuable insights into the underlying mechanisms of genetic diseases and developmental disorders, paving the way for potential therapeutic interventions. Further research is essential to fully elucidate the consequences of chromatin loop disruptions on human health and to explore targeted strategies for mitigating their effects.

In our analysis (Figure 4), we identified 2,582 chromatin loops in the muscular tissue library RT, 268 of which overlapped with 1,523 GWAS SNPs. This overlap is significantly higher compared to 2,582 randomly generated regions, which showed only 391 overlap, showing a high enrichment of GWAS SNPs in the identified loop regions. This significant enrichment emphasizes the functional relevance of chromatin loops in mediating genetic variation and disease susceptibility.

**Nuclear compartmentalization**

Non-coding variants can also disrupt the delicate balance between active (A) and inactive (B) nuclear compartments, which can profoundly affect gene expression and cellular function (Figure 5). These compartments, characterized by distinct chromatin states and transcriptional activities, are essential for maintaining genome stability and regulating proper gene expression [92]. The segregation of these compartments is mechanistically driven by phase separation principles and differential protein-chromatin interactions. Molecular mapping studies have demonstrated that these compartments are not static structures but rather dynamic entities that can shift during cellular differentiation, with genomic regions capable of switching between A and B states. Active compartments are gene-rich and transcriptionally active, while inactive compartments are gene-poor and typically repressed [93, 94].

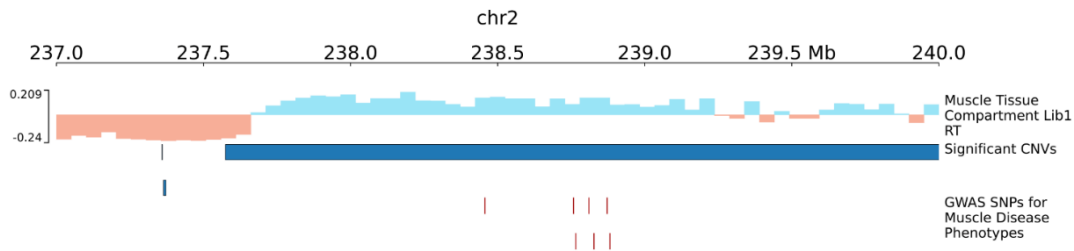

**Figure 5: Overview of compartment generated using Hi-C data.** Compartments A and B analysed from muscle tissue Hi-C data is shown in lightblue for compartment A and lightred for compartment B. The compartment regions are mapped with TAD domains, CNVs and GWAS SNPs.

Recent research has highlighted the significant influence of structural variants on nuclear compartmentalization. Duplications of chromatin regions can cause an overrepresentation of specific compartments, thereby disturbing the regulatory balance and inducing abnormal gene expression patterns. This compartmental dysregulation occurs through perturbation of architectural proteins and altered chromatin compaction states. This is especially apparent in cancer, where amplification of oncogenes within active compartments drives unchecked cell proliferation [95-97]. For instance, Taberlay et al. [98] demonstrated that cancer-associated genome restructuring leads to compartment switching and subsequent changes in gene expression. Similarly, Akdemir et al. [99] showed how structural variations in cancer genomes result in enhancer hijacking and oncogene activation through alterations in 3D genome topology.

On the other hand, deletions within chromatin domains can eliminate crucial regulatory elements, thereby disrupting nuclear architecture and leading to gene misregulation. These deletions can disrupt chromatin folding patterns across multiple scales, from local enhancer-promoter interactions to higher-order compartmental organization, contributing to to various disease phenotypes [100]. CNVs that affect inactive chromatin regions, commonly associated with B compartments, can have substantial consequences. Deletions in these regions can reposition nearby genes into repressive chromatin environments, reducing gene expression and potentially contributing to developmental disorders [101, 102]. Fudenberg and Pollard [103] used polymer simulations to demonstrate how large-scale genomic rearrangements affect chromatin compartmentalization, revealing that compartment disruption follows predictable biophysical principles based on the size and location of structural variants. Additionally, large deletions can disrupt chromatin loops and reorganize nuclear compartments, leading to the collapse of long-range enhancer-promoter interactions and causing widespread changes in gene expression profiles [90].

The integration of high-resolution chromatin interaction maps with CNV data provides valuable insights into how CNVs alter nuclear architecture. This multi-modal approach reveals the hierarchical nature of nuclear organization disruption, from local topology changes to global nuclear architecture perturbation.Understanding these interactions is crucial for predicting the functional impact of CNVs and for developing targeted therapies aimed at restoring proper chromatin organization.

In conclusion, CNVs significantly disrupt nuclear compartmentalization through multiple mechanistic pathways, including altered protein binding, disrupted phase separation, and perturbed chromatin folding patterns, leading to aberrant gene regulation and contributing to a wide range of diseases. Continued research into chromatin interactions and nuclear architecture is essential for advancing our understanding of CNV biology and developing effective interventions for conditions associated with CNVs.

**Disrupting single promoter-enhancer interactions**

Promoter-enhancer interactions are vital for precise gene regulation, where enhancers loop to contact promoters and enhance transcriptional activity. Genetic variants, such as CNVs and GWAS SNPs that either delete enhancers or insert disruptive sequences between an enhancer and its target promoter can impair these crucial interactions, leading to altered or aberrant gene expression. For instance, the deletion of an enhancer can silence genes essential for cell differentiation, potentially resulting in developmental disorders or diseases such as cancer [104]. The mechanistic disruption often involves perturbation of transcription factor binding sites and altered chromatin accessibility states. Quantitative analysis of interaction frequencies shows that enhancer deletions can reduce promoter contacts compared to wild-type configurations [105].

One study demonstrated that a CNV deleting an enhancer upstream of the *SOX9* gene disrupted its interaction with the promoter, leading to aberrant gene expression and skeletal malformations [106, 107]. This finding underscores the essential role of enhancers in maintaining normal gene expression and highlights the severe phenotypic consequences that result from their disruption. Interestingly, research by Long et al. [108] revealed that enhancer redundancy can sometimes buffer against such deletions, adding complexity to the understanding of enhancer function in both development and disease.

Similarly, studies have shown that the *MYC* oncogene, overexpressed in 50-60% of cancers, is tightly regulated by long-range interactions with non-coding super-enhancers. These super-enhancers form specialized regulatory domains characterized by high levels of H3K27ac and dense clusters of transcription factor binding sites [109, 110]. Research has demonstrated that *MYC* overexpression strengthens these enhancer-promoter interactions, increasing chromatin contact frequency and driving aberrant gene expression [110, 111]. This process involves both structural changes in chromatin organization and alterations in the local epigenetic landscape [110, 111]. Disruptions in chromatin architecture, such as duplications that amplify these interactions or deletions that weaken them, significantly affect *MYC* activity and contribute to oncogenesis. Recent CRISPR/Cas9 studies targeting these super-enhancer elements showed marked reduction in MYC expression in cancer cells, highlighting promising therapeutic applications [112].

These studies emphasize the importance of understanding disruptions in promoter-enhancer interactions to unravel the mechanisms underlying genetic diseases and to develop targeted therapies, especially with the integration of chromatin interactions with epigenetic marks, depicted as an example in Figure 6. High-resolution chromatin interaction maps, coupled with advanced genomic tools like CRISPR/Cas9, have been pivotal in validating the functional consequences of CNVs. The development of innovative methods such as HiChIP [113] and Micro-C [114] has further improved our ability to map chromatin interactions with unprecedented resolution, providing deeper insights into the regulatory landscape affected by genetic variants.

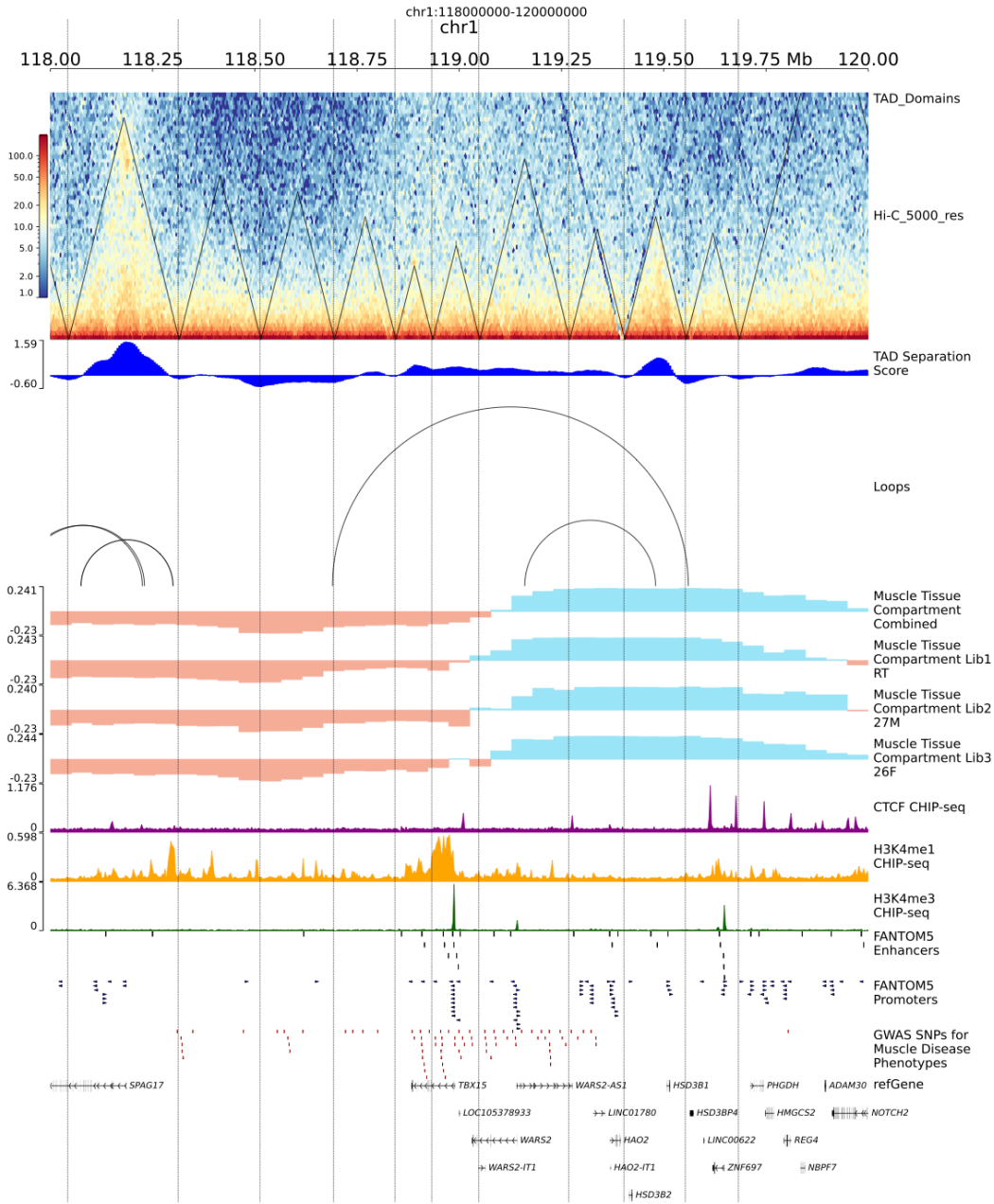

**Figure 6:** A comprehensive plot of a significant chromosome region. From top to bottom showing TAD domains, TAD boundaries, Hi-C data, TAD separation score, Loops, Compartment Analysis for combined library and separated libraries (RT, 27M, 26F), CTCF CHIP-seq, H3K4me1 CHIP-seq, FANTOM enhancers and promoters signal, GWAS SNPs for muscle disease phenotypes and refGene.

**Disrupting super-enhancers**

Super-enhancers are clusters of enhancers that drive the robust expression of genes critical for maintaining cell identity and function. Disruptions to these super-enhancers by CNVs can lead to substantial changes in gene expression, thereby affecting cellular processes and contributing to disease pathogenesis. The integrity and spatial organization of super-enhancers are essential for their regulatory function, and CNVs that delete, duplicate, or rearrange super-enhancer regions can profoundly alter the transcriptional landscape, with significant consequences for gene regulation.

For example, Yi et al. [110]. demonstrated that deletions within super-enhancers associated with the *MYC* oncogene resulted in reduced *MYC* expression, impairing cell proliferation and tumorigenesis in colorectal cancer models. This study underscores the importance of maintaining super-enhancer structural integrity for proper oncogene regulation and highlights how CNV-induced disruptions can directly influence gene expression. Further investigations revealed that specific super-enhancer elements exhibit distinct temporal activation patterns during tumor progression, with early-activated elements being crucial for tumor initiation while late-activated elements drive metastasis [109, 112]. Supporting these, Lovén et al. [115] showed that cancer cells are highly dependent on oncogenes regulated by super-enhancers, identifying these elements as potential therapeutic targets. The field has advanced to show that super-enhancer-associated genes show differential sensitivity to targeted disruption based on their chromatin architecture.

In autoimmune disorders, super-enhancers disruptions can alter the expression of immune-related genes. Dimitri et al. [116] demonstrated that super-enhancer disruptions at the IL2RA locus play a critical role in immune regulation and the development of autoimmunity. Using CRISPR-based mapping of *IL2RA* enhancers, they identified key enhancers whose disruption significantly altered immune cell function and increased disease risk [116]. Single-cell chromatin profiling has now uncovered distinct super-enhancer states across immune cell subpopulations, with disease-associated variants preferentially affecting specific cellular contexts [117]. Remarkably, deleting a disease-associated enhancer within the *IL2RA* super-enhancer protected mice from autoimmune diabetes, emphasizing the role of super-enhancer integrity in immunity and disease susceptibility. This aligns with Vahedi et al. [118], who demonstrated that super-enhancers are key regulators of inflammatory gene programs in T cells, further emphasizing their critical role in immune function and autoimmune diseases.

Understanding the vulnerability of super-enhancers to CNV disruptions holds significant therapeutic potential. Potential treatments include CRISPR-based genomic editing or small-molecule therapies aimed at restoring disrupted super-enhancer activity, as discussed by Xiang-Ping et al. [119]. Investigating how CNVs affect super-enhancers is crucial for unraveling the regulatory mechanisms of gene expression and may lead to the development of novel diagnostic and therapeutic strategies. The development of selective super-enhancer modulators delivered through nanoparticle-based systems has demonstrated therapeutic efficacy in animal models of cancer and autoimmune disease.

**Challenges and future directions**

The findings from this study underscore the pivotal role of non-coding CNVs in regulating gene expression and disrupting regulatory regions across the human genome. A major challenge in interpreting non-coding variants arises from the limitations of standard variant annotation and curation processes. These processes are typically designed to evaluate whether variants are benign or pathogenic, following established guidelines such as those from ClinGen and College of Medical Genetics and Genomics (ACMG). However, these frameworks primarily address non-synonymous variants in exonic regions, leaving non-coding CNVs, especially those without clear gene associations, often underexplored and lacking functional annotations in databases.

Our results illustrate that non-coding CNVs, particularly those located in intronic or intergenic regions, can have substantial functional impact. Those impact includes disruption of cellular functions, gene regulation, transcription factors binding activities and signalling pathways. Given that non-coding regions account for approximately 98% of the human genome, these variants offer significant insights into complex diseases, such as cancer and neurocognitive disorders. Expanding research into non-coding CNVs is essential for understanding their role in genetic disorders and improving predictive models. The study also highlights the importance of incorporating Hi-C data in the investigation of non-coding regions. Chromatin interaction maps offer a three-dimensional perspective on the spatial organization of the genome, revealing that non-coding variants may influence gene regulation more profoundly than SNVs due to their broader-scale alterations. To achieve this, existing tools such as DeepGenePrior [120] and CNVDeep [121] need to adapt their strategies for identifying CNV associations. This includes incorporating chromatin interactions to prioritize CNV regions, particularly non-coding CNV regions. This insight emphasizes the potential of targeting enhancer regions and understanding chromatin interactions to impact gene expression regulation. Despite the advancements provided by Hi-C, there remain challenges in integrating these data with genomic and clinical information. Future research should focus on developing more refined in vivo models to validate chromatin interactions and their regulatory consequences. Such studies could pave the way for innovative gene-based therapies and advance precision medicine, enabling personalized treatment strategies that account for individual genomic profiles.

Furthermore, exploring the impact of structural variants on topologically associating domains (TADs), chromatin loops, and super-enhancers presents significant challenges and opportunities. Detailed characterization of overlap between TAD boundaries, chromatin loops and GWAS-asscociated SNPs could protray distinct patterns of evolutionary conservation and tissue-specific regulatory activity. Notably, loops containing disease-associated variants significantly higher sequence conservation across vertebrates and enrichment for muscle-specific transcription factor binding sites. Disruptions in TAD boundaries and chromatin loops can lead to widespread gene misregulation and disease phenotypes. Future research should aim to elucidate the mechanisms by which CNVs and other structural variants affect these crucial elements of chromatin organization. This includes developing novel techniques to map and analyze chromatin interactions with greater resolution and accuracy, which will be critical for understanding the full scope of their effects.

Addressing these challenges will require interdisciplinary approaches, combining advanced genomic technologies, computational models, and functional assays. Collaborative efforts across research domains will be essential to advance our understanding of non-coding variants and their role in gene regulation and disease. Continued innovation in these areas promises to enhance our ability to diagnose, predict, and treat complex genetic disorders, ultimately leading to more effective and personalized therapeutic interventions.

**Funding**

YL and SA are supported by Australian Government Research Training Program Scholarship (RTP).

**Acknowledgements**

NA

**Author contributions**

YL designed the study. YL wrote the manuscript with help from HAR, NF and SA. YL carried out all the analyses including the statistical analyses, region identification, text mining, and Hi-C data analyses. YL created all the figures and tables. All authors have read and approved of the final version of the paper.

**Conflicts of Interest**

The authors declare no competing financial/non-financial interests.